\documentclass[
aip,
amsmath,amssymb,
groupedaddress,
reprint,
]{revtex4-1}

\usepackage{graphicx}
\usepackage{dcolumn}
\usepackage{bm}

\usepackage{textcomp}
\usepackage{float}
\usepackage{helvet}
\usepackage{hyperref}
\usepackage{url}
\usepackage{xcolor}
\usepackage[utf8]{inputenc}
\usepackage[T1]{fontenc}
\usepackage{mathptmx}
\usepackage[normalem]{ulem}
\usepackage{varwidth}
\usepackage{titlesec}

\newcommand*{\myfont}{\fontfamily{phv}\selectfont}

 \titleformat{\section}
  {\myfont\fontsize{10}{13}\bfseries}{\thesection}{1em}{}

 \titleformat{\subsection}
  {\myfont\fontsize{9.5}{11.5}\bfseries}{\thesubsection}{1em}{}

\begin{document}

\preprint{AIP/123-QED}

\title{Integrated table-top facility for the study of Whispering Gallery Modes in dynamic liquid micro-cavities coupled to sub-micron tapered fibers \vspace{0.5 em}}

\author{Meenakshi Gaira }
\email{meenakshi.gaira@tifr.res.in}
\author{C.S. Unnikrishnan}
\affiliation{Tata Institute of Fundamental Research, Mumbai 400005, India
}

\date{\today}

\begin{abstract}

\vspace{1.0 em}
A complete integrated table-top facility for the study of high-$Q$ Whispering Gallery Modes in solid and liquid micro-cavities is described, with emphasis on the in situ fabrication of reliable tapered fibers of sub-micron waist sizes for coupling light into time-dependent liquid micro-cavities. The experimental parameters have been chosen to get nearly adiabatic tapers, with their waist size consistent with the theoretical model. The oscillations in the transmitted power during fiber tapering are monitored to check the coupling of higher order modes and identify the point when the fiber is suitable for pure single mode coupling. The fabricated tapered fibers have greater than 85\% transmission on the average and very good polarization fidelity. The tapered fibers have been used for efficiently exciting and detecting WGMs of $Q\sim10^7$ in silica microspheres and of $Q\  \text{up to}\ 7 \times 10^7$ in microdrops of silicone oils.

\end{abstract}

\maketitle

\section{\label{sec:level1}INTRODUCTION}

Whispering Gallery Modes (WGMs) of micro-cavities have applications in non-linear optics \cite{Kerry2, Beckmann}, optomechanics \cite{Kippenberg}, cavity QED \cite{Kerry1, Kimble}, biosensing \cite{Arnold,Wu}, etc.~(see Ref. \onlinecite{Ilchenko, Kerry1}) because of their high Quality factor ( $Q$ ) \cite{Kerry1, Henriet} and confinement of high optical power in very small mode volume. Tapered optical fibers \cite{Cai, Lin, Riviere} are the most commonly used devices for coupling light into the microsphere cavities,  as compared to other couplers like prisms\cite{Braginsky, Klitzing}, eroded fibers \cite{Dubreuil} and angle polished fibers \cite{Vladimir} because they are efficient, compact and they filter out the higher order modes which otherwise add noise in the main signal in the fundamental mode. Tapered fibers are also used in fiber couplers, fiber splitters, sensors \cite{Polynkin, Villatoro, Zhang, Herrera}, for coupling light into photonic crystals \cite{Sadgrove, Thompson}, in  trapping and detecting ultra-cold atoms\cite{Vetsch, Nayak, Morrissey}, etc.~(see Ref. \onlinecite{Lou}).

Tapered fibers are widely used for efficiently and selectively exciting WGMs in solid micro-cavities. However, there are only a few reported works\cite{Dahan, Jonas, Hossein, Kaminsky} on using tapered fibers for coupling light to liquid micro-cavities because of their dynamic geometry that keeps changing fast with time. Liquid microdrops are time-dependent micro-cavities with naturally very smooth surface as compared to artificially made solid micro-cavities. Unlike solid micro-cavities, these cavities are dynamic as many physical processes, like evaporation and thermal oscillations along with mechanical deformations and oscillations, continuously take place depending on the external surrounding conditions. Optical WGMs in liquid microdrops provide a novel probe for better understanding of dynamics of liquid drops (like their capillary\cite{Maayani} and acoustic modes\cite{Dahan}, fluctuations in evaporation rate, etc.), properties of liquids (like bulk modulus, surface tension, absorption, viscosity, etc.), and provide a platform for investigating micro-optomechanics\cite{Dahan, Maayani, Jonas}. Also, the measured optical $Q$ can be used to estimate a reliable upper limit of the absorption coefficient of liquids. This could be advantageous for liquids with very low absorption $\lesssim 10^{-3}~\rm{cm}^{-1}$, for which determining absorption becomes difficult when using standard methods (which use bulk samples) due to systematic errors of the same order.\cite{Gaira_conf}

Here, we report the development of a complete integrated table-top facility for the study of high-$Q$ WGMs in solid and liquid micro-cavities. Our motivation for building this set up is to experiment with taper-fiber coupled liquid microdrops to explore these physics-rich systems for novel studies and applications. The facility consists of four main units --- set up for fabrication and characterization of tapered fibers, set up for fabrication of silica microspheres and formation of liquid microdrops, set up for exciting and detecting WGMs in micro-cavities using tapered fibers, and a vibration-isolated experimental platform inside a laminar flow enclosure unit for protecting the tapered fibers and the micro-cavities from dust particles during experiments. We have compiled in this self-contained paper the technical details of our experimental facility along with the required information from the literature, for completeness.

This paper is organized as follows. In section 2, we describe the salient features of all the units comprising the facility. In section 3, we compare the tapering process with its theoretical model. In sections 4 and 5, we explain the propagation of light through a tapered fiber to understand how to fabricate tapered fibers which have high transmission and which can completely filter out the unwanted higher order modes. We also discuss the analysis of transmission data of a tapered fiber, taken while tapering. In section 6, we briefly describe an experiment to check whether the cylindrical symmetry of the fibers is maintained while tapering, by comparing their polarization response before and after tapering. In section 7, we show the observations of high-$Q$ WGMs in silica microspheres. In section 8, we conclude with the measurement of the exceptionally high $Q$ values of WGMs in  microdrops of silicone oils, using cavity ring-down interferometry\cite{Rasoloniaina, Dumeige, Ye}.

\section{\label{sec:level2}EXPERIMENTAL SETUP AND PROCEDURE}

\subsection{\label{subsec:level1}Fabrication of solid microspheres and formation of liquid microdrops}

\begin{figure}[t]
\includegraphics[width=1.0\linewidth]{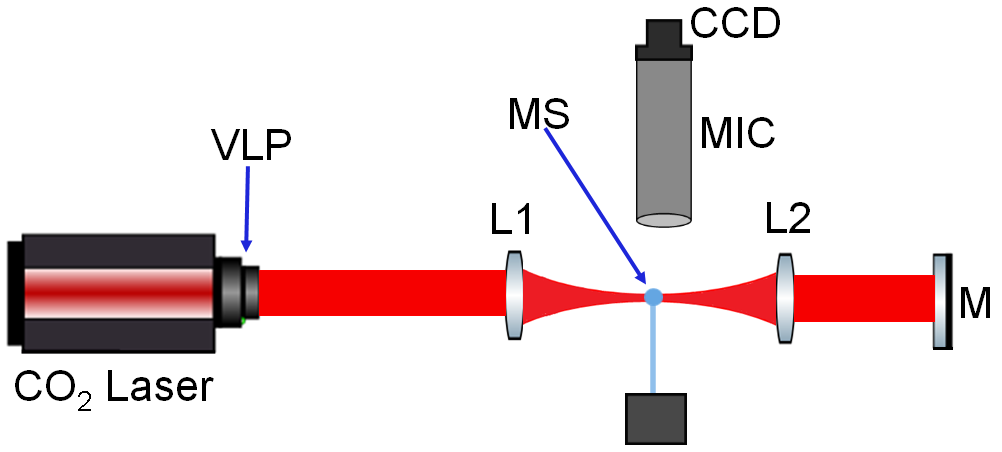} 
\caption{Schematic of experimental set up for fabrication of silica microspheres. L1 and L2: lenses, M: mirror, MS: silica fiber's tip which has got re-formed into a microsphere, MIC: microscope with a CCD camera attached to it, and VLP: visible laser pointer used for optical alignment of $\rm{CO_2}$ laser beam. L1, L2 and MS are on XYZ mounts or stages for alignments. \vspace{5 mm}}
\label{fms}
\end{figure}

\begin{figure}[t]
 \centering
\includegraphics[width=0.7\linewidth]{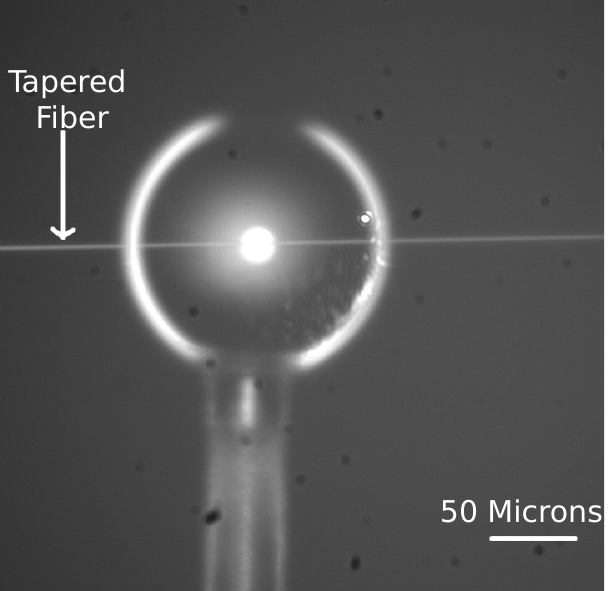} 
\caption{Microscope image of a silica microsphere of diameter $\approx 145$~{\textmu}m, kept very close (sub-wavelength gap) to a tapered fiber to excite its WGMs.}
\label{mic_tap}
\end{figure}

We study WGMs in two kinds of micro-cavities --- solid microspheres and liquid microdrops. We fabricate solid microspheres by melting the tips of silica fibers using two balanced counter-propagating  $\rm{CO_2}$ laser beams, generated by retro-reflection of a beam by a Si mirror. Surface tension re-forms the fiber tips to spheres. A schematic of the set up is shown in Fig. \ref{fms}. The balanced beams avoid asymmetric radiation pressure which otherwise produces bent or asymmetric (about the fiber stem) microspheres. There are two ZnSe lenses for focusing the beams and the fabrication process is monitored with a microscope and a CCD camera attached to it. Since $\rm{CO_2}$ laser beam ($10.6$~{\textmu}m wavelength) is invisible, a visible laser pointer, which consists of a diode laser ($650$~nm wavelength) and a ZnSe beam combiner/splitter, has been mounted on the front panel of the $\rm{CO_2}$ laser. Its beam is co-aligned with the $\rm{CO_2}$ laser beam for optical alignments. We produce microspheres with diameters in the range 30--300~{\textmu}m with this setup (see Fig. \ref{mic_tap}).

\begin{figure}[t]
 \centering
\includegraphics[width=0.8\linewidth]{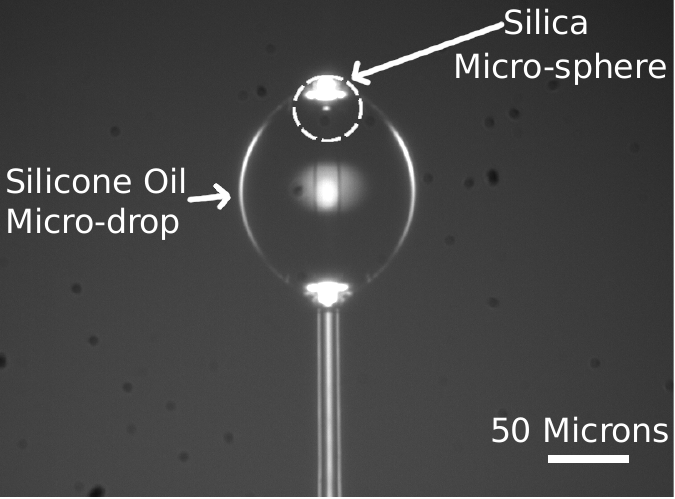} 
\caption{Microscope image of a silicone oil microdrop of diameter $\approx 105$~{\textmu}m around a silica microsphere of diameter $\approx 40$~{\textmu}m (bordered by a dashed circle). We form such microdrops by dipping silica microspheres in silicone oils. Note that the centroid of the liquid microdrop is below the centroid of the much smaller solid microsphere.}
\label{sil_drop}
\end{figure}

We form a liquid microdrop (see Fig. \ref{sil_drop}) by dipping a small silica microsphere inside the liquid. A microdrop gets formed around the solid microsphere due to cohesive and adhesive forces in and between the liquid and silica. Microdrops of silicone oils with diameter 50-200~{\textmu}m are used in our experiments. Silicone oils have been chosen for the microdrops because it is technically easier to excite and detect WGMs in these drops. This is due to the following reasons. a) very low evaporation rate, b) refractive index close to silica which helps in phase matching,\cite{Lin} c) very low optical absorption, and d) good tendency to form drops around silica microspheres which allows easy positioning and alignment of the drops. (this tendency depends on the viscosity, the cohesive forces in the liquid and the adhesive forces between the liquid and silica).

\begin{figure}[t]
\includegraphics[width=1.0\linewidth]{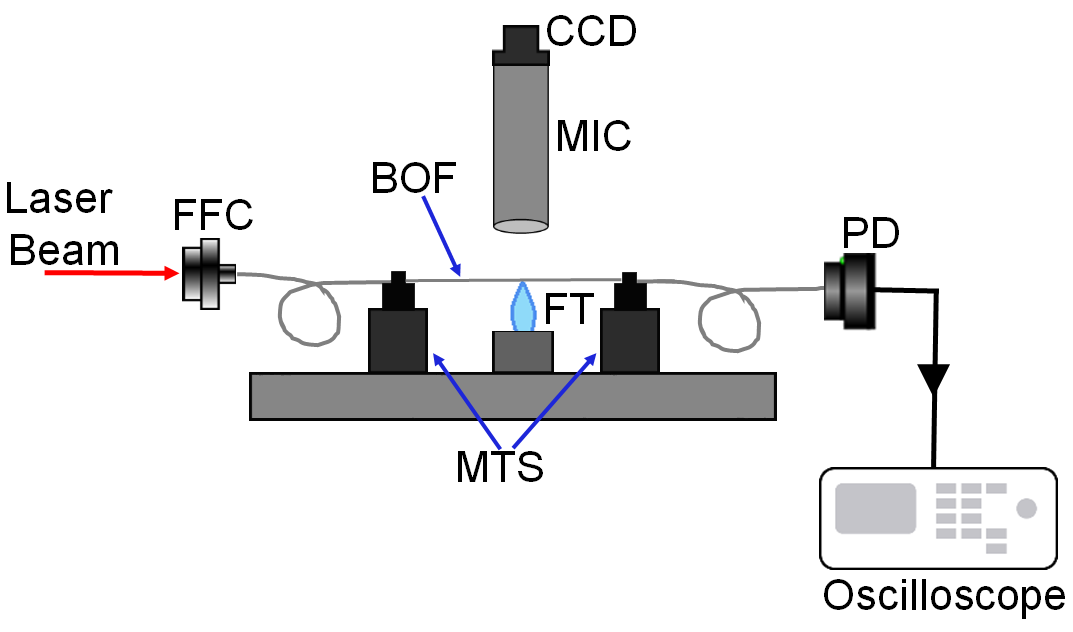} 
\caption{Schematic of experimental set up for fabrication of tapered fibers. MTS denotes two identical motorized translation stages. BOF: bare optical fiber, FT: flame-torch (hydrogen-oxygen torch), FFC: free space fiber coupler, PD: photo-detector, and MIC: microscope with a CCD camera attached to it. }
\label{ftfs}
\end{figure}

\subsection{\label{subsec:level2}Fabrication of tapered fibers}

For efficiently and selectively exciting (and detecting) WGMs, approximately adiabatic \cite{Love, Ravets, Ward_review} tapers are needed. A tapered fiber is called adiabatic if the taper angle is small enough everywhere such that negligible power couples from the fundamental mode to the higher order modes and the radiation modes (which cause power loss), as it propagates along the taper. For adiabatic tapers of single mode fibers, the `local mode theory' \cite{Love} says that at every local point of the tapered fiber, the field distribution and propagation constant can be described by the fundamental mode of a cylindrical guide with the same geometry as the local point. So, during propagation, the fundamental mode of one local point transforms to the fundamental mode of successive local point without any loss of power.

Heat and pull is the most commonly used method for fabricating tapered fibers of sub-micron waist sizes.\cite{Orucevic, Anthony, Hauer, Ward, Ward_review} A heat source like $\rm{CO_2}$ laser, flame torch or micro-furnace softens a fiber and simultaneously two motorized translation stages pull the fiber from both the sides. The advantages and disadvantages of each of these heat sources are mentioned in Ref. \onlinecite{Ward_review}. We use a hydrogen-oxygen torch as the heat source. A schematic of the set up is shown in Fig. \ref{ftfs}. A hydrogen-oxygen torch is preferred over other gas torches because it has a clean flame and leaves no residue other than water vapor, which also disperses in the air leaving a clean taper. This helps in making tapered fibers with good performance. Single mode 780 HP fiber from Thorlabs, with $4.4$~{\textmu}m core diameter and $125$~{\textmu}m cladding diameter, is used in our experiments. A long working distance microscope (from Navitar) with a CCD camera is mounted on one side for monitoring the whole fabrication process. 

The first step in the fabrication of a tapered fiber is splicing a bare optical fiber with an optical patch cable, which is connected to a free space fiber coupler that couples the laser light (Toptica DL100 ECDL at 780 nm) into the fiber. The other end is connected to a photo-detector using Thorlabs BFT (Bare Fiber Terminator) connector which is like a temporary FC connector for a bare fiber. Next, the protective coating around the fiber is stripped off over 2 cm near the center of the bare optical fiber. Then the stripped part is cleaned with lint-free wipes soaked in alcohol. Subsequently, the fiber is clamped on the two motorized stages using fiber clamps such that the stripped part comes between the two clamps. After this, the hydrogen-oxygen torch is moved below the stripped part of the fiber with an air gap of nearly 3-4 mm between the nozzle and the fiber. In our case, the nozzle diameter is 1mm. The hydrogen gas flow is set between 25 to 50 ml/min and the flame uses oxygen from the surrounding atmosphere. All these parameter values are chosen to get a sufficiently wide and short flame which provides a wide hot zone (part of the flame which softens the fiber) with low gas pressure and nearly uniform temperature along its length. Such kind of hot zone is required for getting adiabatic tapers. This will be discussed in detail in the next sections. With these settings, 1 to 4 mm hot zone length can be obtained, depending on the position of the flame with respect to the fiber.

\begin{figure}[t]
\includegraphics[width=1.0\linewidth]{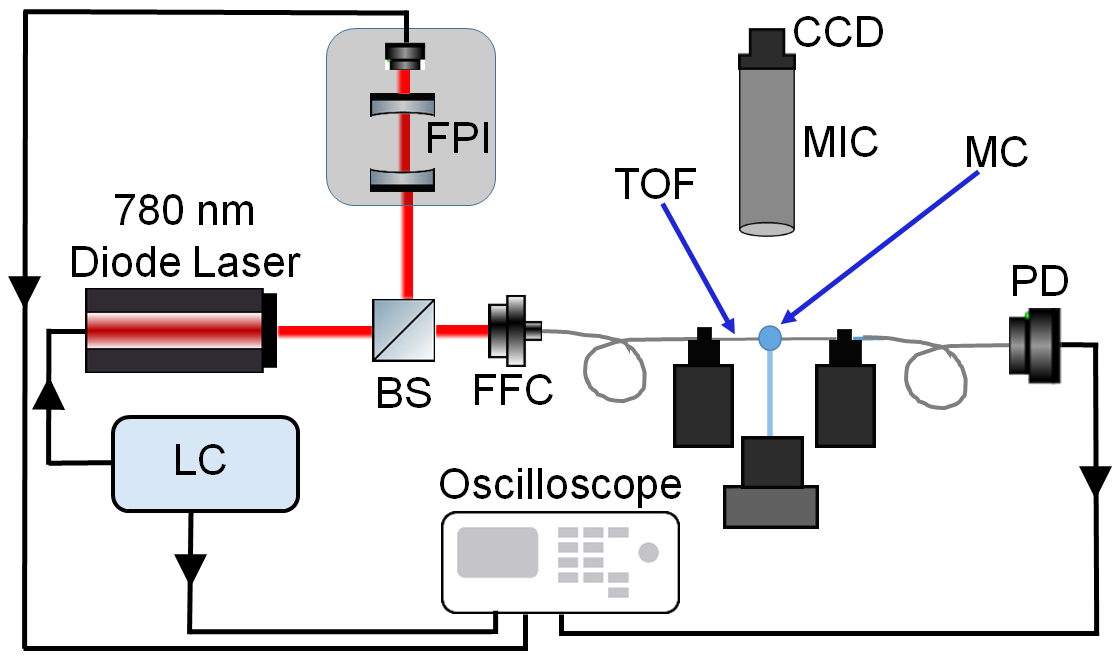} 
\caption{Schematic of experimental set up for exciting and detecting WGMs in micro-cavities. MC denotes a micro-cavity mounted on a 3-Axis nano-positioning stage and a goniometer stage. Other elements are TOF: tapered optical fiber, BS: beam splitter, FPI: Fabry-P{\'{e}}rot interferometer, FFC: free space fiber coupler, PD: photo-detector, LC: laser controller, and MIC: microscope with a CCD camera attached to it. Black lines represent co-axial signal cables.}
\label{ews}
\end{figure}

\begin{figure*}[t]
\includegraphics[width=1.0\linewidth]{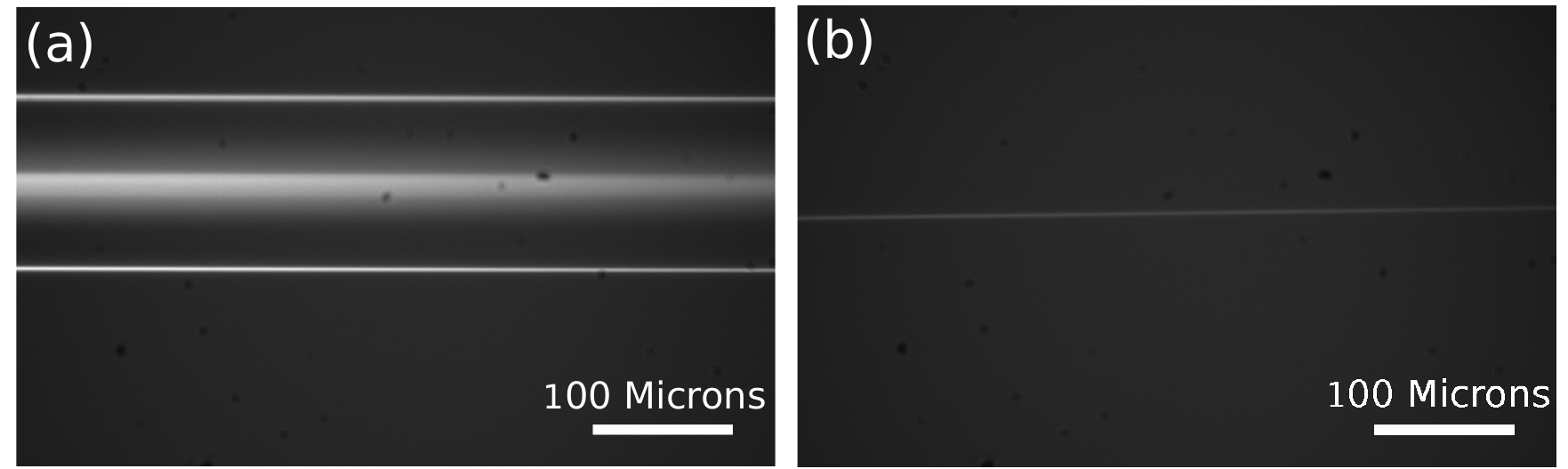} 
\caption{Microscope images of a fiber (a) before tapering, diameter $\approx$ 125~{\textmu}m, (b) after tapering, diameter $<$ 1~{\textmu}m.}
\label{ftm}
\end{figure*}

As the torch is moved close to the fiber, the part of the fiber under the flame starts glowing and then immediately fiber pulling is started using the two computer-controlled motorized stages. The pulling speed is not very critical for obtaining good quality tapers. Prior works report pulling speed values ranging from $20$~{\textmu}m/s to $200$~{\textmu}m/s.\cite{Orucevic, Ward_review, Ravets} We chose a slower speed of $20$~{\textmu}m/s for each motorized stage to have a longer time span between the single mode operating point and the breaking point of the fiber (explained in Sec. 5). This allows us to have better control on stopping the tapering process. The stages are stopped immediately after getting the desired taper waist at which almost all higher order modes get filtered out leaving only the fundamental mode. This stopping point is decided by observing the transmission through the fiber while tapering. For this, a photo-detector and an oscilloscope are used. This will be discussed in detail in the following sections. Immediately after stopping the stages, the torch is moved away from the fiber. Fig. \ref{ftm} shows the microscope images of a fiber before and after tapering.

\subsection{\label{subsec:level3}Exciting and detecting WGMs in micro-cavities}

After fabrication of a tapered fiber, micro-cavities are brought closer to the tapered fiber to excite their WGMs. A schematic of the set up for coupling light into these cavities is shown in Fig. \ref{ews}. Half of the set up is the same that is for the fabrication of tapered fibers. Before the fiber coupler, the laser light is split into two parts using a 70:30 cubic beam splitter. The major part of the light is coupled to the tapered fiber and the small part goes to a Fabry-P{\'{e}}rot interferometer (with a detector attached to it) which provides a relative frequency scale. Two microscopes are used for monitoring the side view and the top view (not shown in the schematic). Both side view and top view are important for aligning the position and orientation of the micro-cavity with respect to the tapered fiber. For alignment, the micro-cavity is mounted on a 3-Axis piezoelectric nano-positioning stage (PI P611.3S) and a goniometer (Thorlabs GNL20).

The laser frequency is scanned to excite the WGMs at the resonance frequencies of the micro-cavities. Two different lasers were used for the observations reported in this paper. Toptica DL100 (ECDL) was used for experiments with solid micro-cavities. It has $\sim 1$~MHz linewidth, which is good enough to resolve resonances with $Q < 10^8$, when using the method of `linewidth measurement' for determining the $Q$ values. Its frequency is scanned by driving the piezo attached to the grating inside it. For liquid microdrops, we use `cavity ring-down' technique (explained in Sec. 8) to measure the $Q$ values. This technique requires very fast frequency scanning. For this, Sacher TEC 55 (diode laser with VHG-based external cavity) at 780~nm wavelength is used . Its frequency is scanned by modulating the injection current using a ramp signal generator. In our experiments, its frequency has been scanned up to $10^6$~GHz/s speeds.

All the data is acquired and saved using an oscilloscope with up to 4~GSa/s acquisition rate. For cavity ring-down measurements with liquid microdrops, which require very fast data detection, the large area photo-detector at the tapered fiber's output is replaced by a fast and high gain detector (Thorlabs APD 430A with 400~MHz bandwidth). High gain is used because experiments for measuring $Q$ are done with low power $<20$~{\textmu}W to minimize thermal, non-linear and radiation pressure effects. \cite{Carmon, Dahan, Kippenberg, Kippenbergkerr, Imad}
\vspace{4em}

\begin{figure}[b]
\centering 
\includegraphics[width=0.9\linewidth]{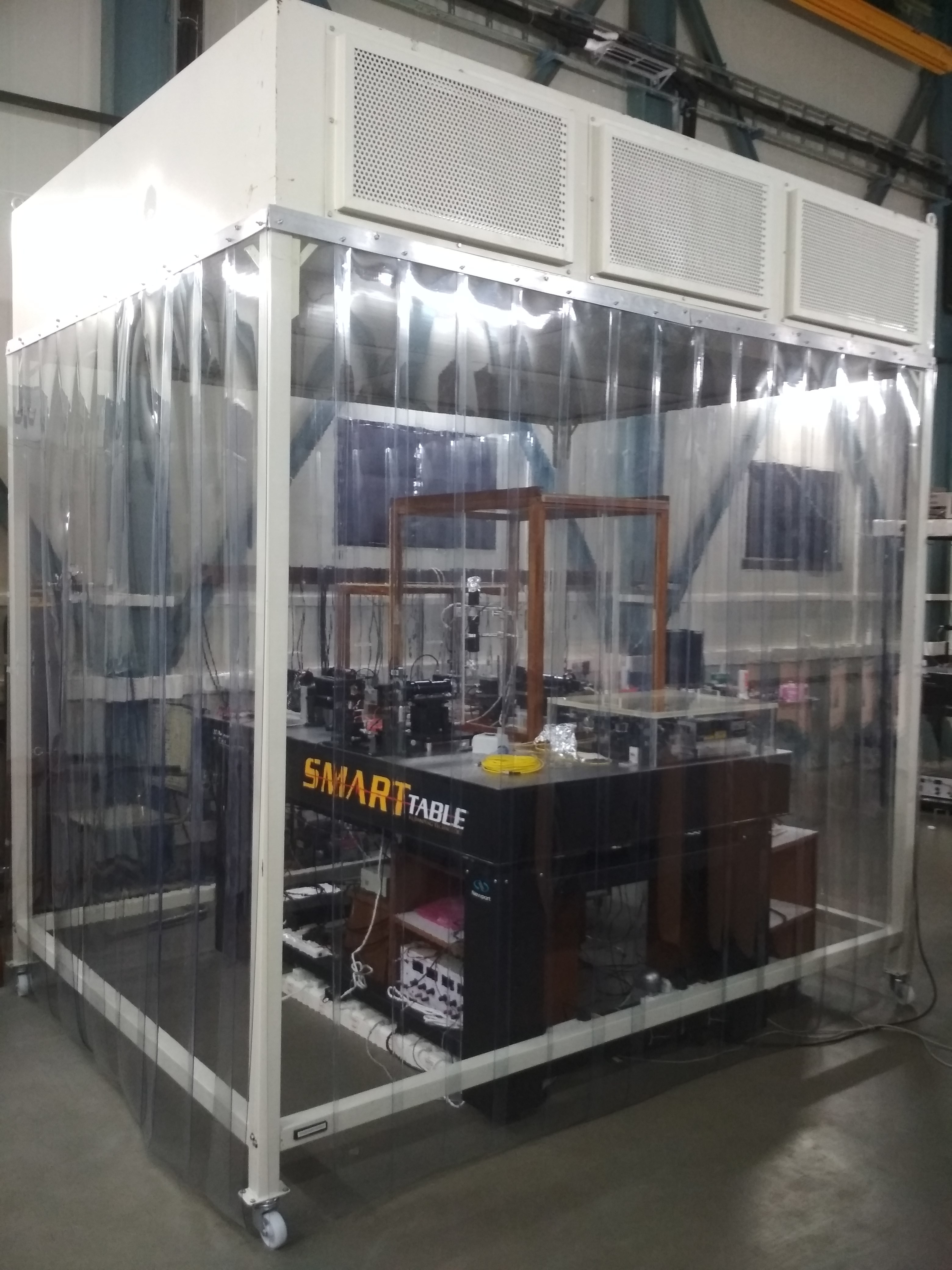} 
\caption{ Laminar flow assembly (dimensions $2.5~\rm{m}\times 1.9~\rm{m} \times2.9~\rm{m}$) with the whole set up covered inside it.  }
\label{lp}
\end{figure}

\subsection{\label{subsec:level4} Enclosure units and vibration isolation}

The entire fabrication and experimental facility has been built inside a laminar flow assembly (shown in Fig.  \ref{lp})  to protect the tapered fibers and the micro-cavities from dust particles in the laboratory environment. Without this air filtering assembly, we were not able to use a tapered fiber for more than an hour or two as many dust particles accumulate on the waist region. Apart from causing light loss due to evanescent scattering, the dust particles prevent the clog-free placing of the fiber's waist near the microsphere for evanescent coupling of light. Although the optical resolution of the microscope is $\sim 1$~{\textmu}m, dust particles of all sizes, from less than 100~nm to few microns, are visible and identifiable under the microscope as these scatter the light propagating through the tapered fiber. After installation of the assembly, we can use a tapered fiber for 2 to 4 days, depending on the quality of the surrounding air which varies with seasons. This laminar flow assembly was custom designed with  dimensions $2.5~\rm{m}\times 1.9~\rm{m} \times2.9~\rm{m}$. This assembly provides us a cleanroom environment of ISO Class 5 standard, which means $\sim 10,000$ particles per $\rm{m}^3$ of size $\geq 0.3$~{\textmu}m. This kind of assembly is useful in labs where a fully isolated cleanroom can't be made due to some technical or operational reasons. The setups are also surrounded by PVC strip curtains to prevent them from dust. Additionally, on the optical table also, the setups have been fully covered by acrylic covers with wooden frames.

While exciting WGMs in micro-cavities, a sub-wavelength gap is to be maintained between the cavity and the tapered fiber. Therefore, the coupling of evanescent light from the fiber to the cavity is sensitive to vibration noise of even tens of nanometers. For this, all the experimental setups have been built on a single floated optical table. Other than blocking dust particles, the PVC strip curtains also block external air flows due to ACs and dehumidifiers which can cause vibrations. All the cables are clamped on the optical table to damp the vibrations that may get transferred through them. All the instruments, optics and optomechanics are mounted firmly. Thus, the possible sources of vibrations, which could be avoided or damped, were taken care of.

\section{\label{sec:level3}SHAPE AND WAIST SIZE OF TAPERED FIBERS}

\begin{figure}[t]
\centering 
\includegraphics[width=1.0\linewidth]{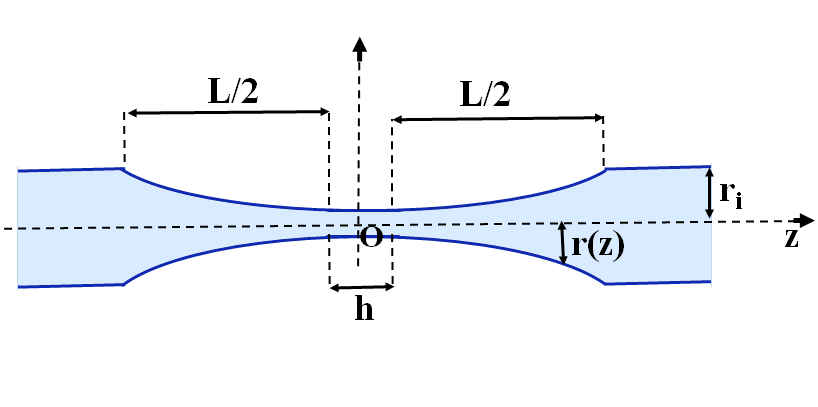} 
\caption{Schematic of the exponential-profile of tapered fibers. $r_i$ is the initial radius, $r_w$ is the waist radius, $h$ is the hot zone length, $L$ is the pulling length, z=0 is marked by $O$.   }
\label{ftf_profile}
\end{figure}

The mathematical profile of a tapered fiber, considering constant and uniform temperature hot zone length, can be obtained using the simple constraint of conservation of mass.\cite{Birks, Orucevic} According to this model, for a stationary flame, the shape of a tapered fiber is exponential (schematically shown in Fig. \ref{ftf_profile}) and is given by

\begin{equation}
r(z) =
\begin{cases}
      r_i \exp \left( \frac{|z|-h/2-L/2}{h} \right) &  \text{for \ \ \ } h/2<|z|<(L+h)/2 \\ \\
      r_i \exp \left( \frac{-L}{2h} \right) & \text{for \ \ \ } 0<|z|<h/2. 
\end{cases}
\label{taper_profile}
\end{equation}

\noindent Here, $r(z)$ is radius of the tapered fiber at axial position $z$, $h$ is the hot zone length, $L$ is the pulling length, $O$ denotes the center of the hot-zone (which is origin for the axial axis), $r_i$ and $r_w$ represent the radius of the initial fiber and the taper waist respectively.

\noindent For pulling time $t$ and constant pulling speed $v$ on both the sides of a fiber, the pulling/tapering length can be written as 
\begin{equation}
L=2vt.
\label{taper_length}
\end{equation}

Using eq. \ref{taper_profile} and eq. \ref{taper_length}, the waist size as a function of pulling time can be written as
\begin{equation}
r_w=r_0 \exp (-vt/h).
\label{fit_equation}
\end{equation}  

\begin{figure}[t]
\centering 
\includegraphics[width=1.0\linewidth]{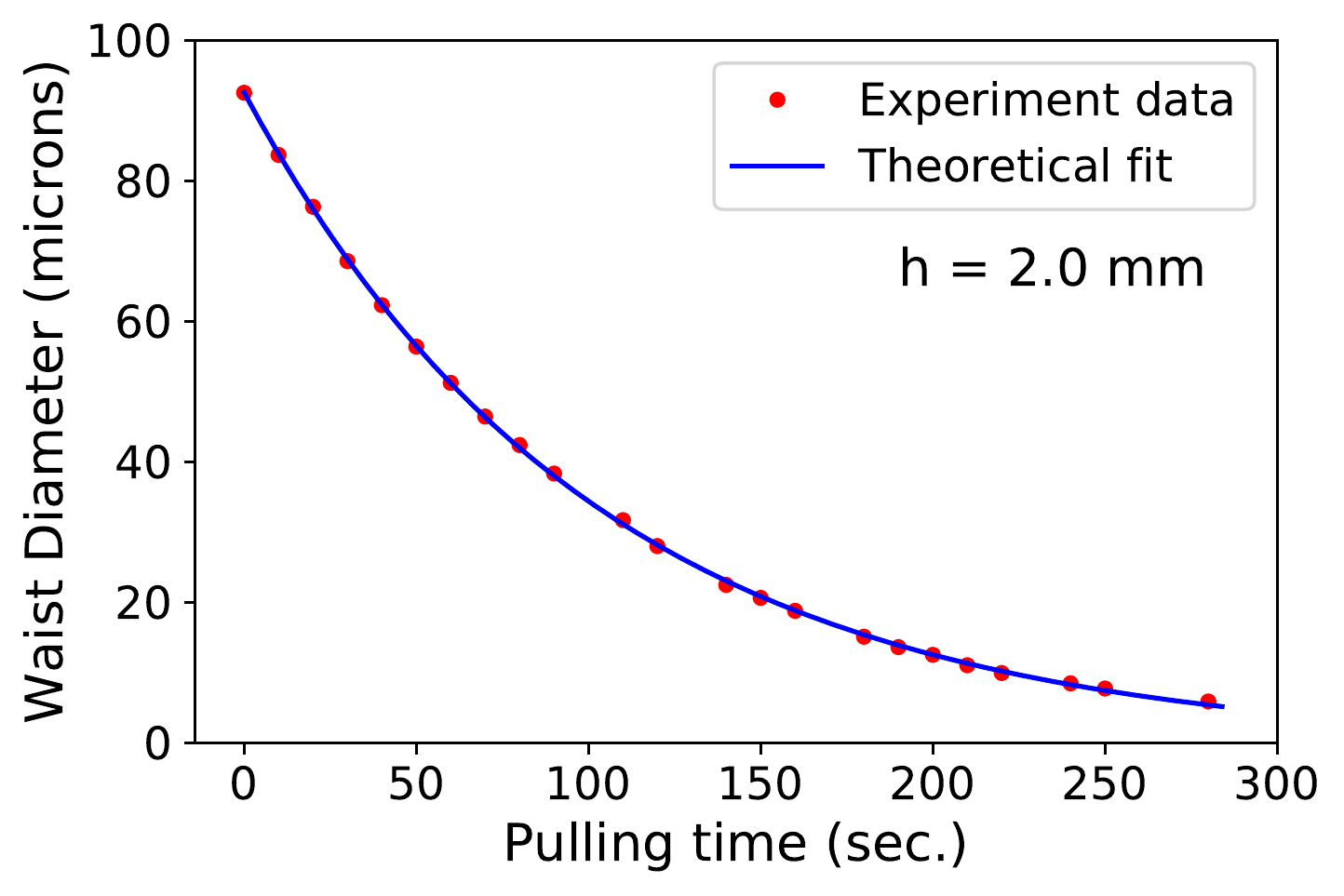} \caption{Change in the waist diameter of a fiber with pulling time while tapering. Red points are the experimental data points and the blue curve is an exponential fit to eq. \ref{fit_equation}. The fitting parameter $h$ for this data was found to be $2.0$ mm.}
\label{expo_plot}
\end{figure}

\begin{figure*}[t]
 \centering 
\includegraphics[width=0.9\linewidth]{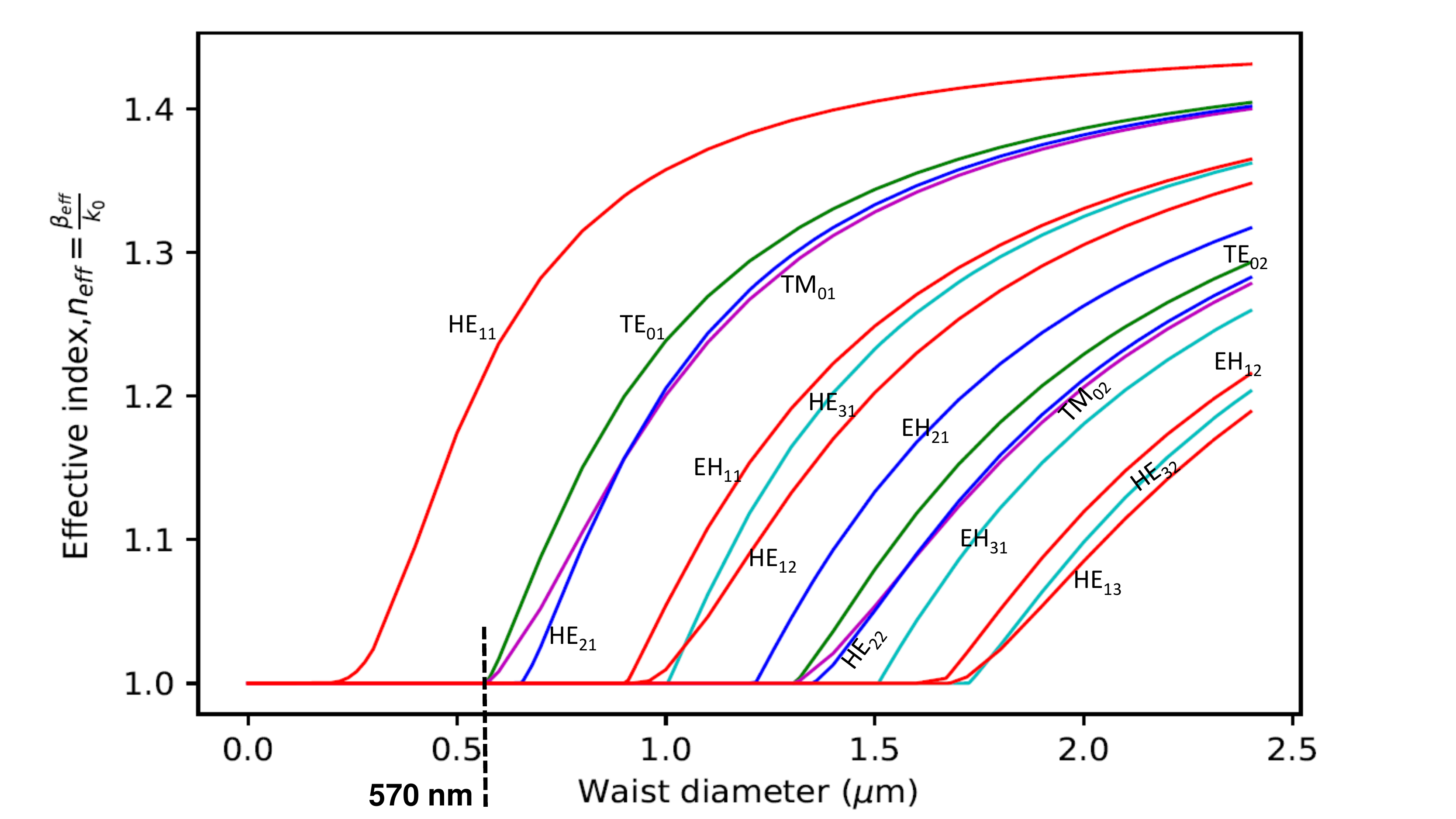} 
\caption{Theoretically calculated effective index for first few modes in a silica-air waveguide as a function of the core diameter for wavelength $\lambda = 780$ nm. The waist of a tapered fiber is a silica-air waveguide with core and cladding refractive index $n_s=1.45$ and $n_a=1$ respectively. The theoretical cut-off waist size, below which the waist can support only a single mode, has been marked by a dashed black line.   }
\label{neff_plot}
 \end{figure*}

A sequence of microscope images of a fiber waist were taken while tapering (using LabVIEW) and then the size of the fiber waist in all the images was measured. The data of change in the waist diameter as a function of time was theoretically fitted with eq. \ref{fit_equation} for multiple tapered fibers. The value of fitting parameter $h$ (which is different for different tapered fibers) was found to be consistent with the range of hot zone lengths (typically 2 to 3 mm) used in the experiments. The fitted plot for one of the tapered fiber is shown in Fig. \ref{expo_plot}. While tapering, the fiber waist moves and oscillates many times due to the flame pressure. Although small height flame is used to minimize this pressure effect, we still had to keep moving the microscope manually while tapering to bring the fiber waist under focus. Many images that were out of focus were not used for the final data. For this reason, the data points in Fig. \ref{expo_plot} are not at equal time intervals. Also, the images after 5~micron waist size were not used because those images were too fuzzy to measure the waist size accurately.

\section{\label{sec:level4}PERFORMANCE OF TAPERED FIBERS}

As light propagates along a tapered fiber, it passes through different zones of optical confinement. At the starting part of the tapered section, the core and cladding are sufficiently thick, therefore, the light is totally confined in them and this core-cladding region supports only a single mode. A gradual transition occurs from this as the thickness decreases and the light sees a core-cladding-air waveguide. This region can also support some higher order modes. Here, the single mode of light coming from one end of the fiber can couple to higher order modes, depending on the slope of the taper. Approximately adiabatic taper \cite{Orucevic, Love, Ravets} minimizes the coupling from the fundamental mode to the higher order modes and also dissipative radiation modes. We can see from eq. \ref{taper_profile} that the slope of the exponential profile depends on the hot zone length. Wider hot zone length gives a more gradual slope. However, too wide hot zone may affect the uniformity of temperature along its length, depending on the nozzle design.\cite{Ward_review, Orucevic, Lin} Thus, we choose a nozzle size (from the available ones), the hydrogen gas flow rate and the position of the fiber with respect to the flame such that we get a sufficiently wide hot zone with nearly uniform temperature along its length to get nearly adiabatic tapers. In our experiments, the hot zone length typically lies between 2 to 3 mm.

Due to small departures from adiabaticity, higher order modes get excited while the fundamental mode propagates from one end of the taper towards the taper waist and then also re-couple to the fundamental mode while propagating from the taper waist to the other end of the taper. These higher order modes contribute to the noise in the signal. Due to axial symmetry of the tapered fiber, the fundamental $LP_{01}$($HE_{11}$) mode can only couple to the higher order modes with the same azimuthal symmetry i.e. $LP_{0m}$($HE_{1m}$). Therefore, probability of coupling to $LP_{02}$($HE_{12}$) mode is the most. But, if the tapered fiber is bent or due to some imperfections in the tapering process, other closer higher order modes with different azimuthal symmetries can also couple, like $TE_{01}$, $HE_{21}$, $EH_{11}$, etc., depending on their coupling efficiencies. \cite{Orucevic,Ravets} During our fabrication process, tapered fibers sometimes slightly bend upwards due to the flame pressure (which however doesn't affect the transmission much). This may also cause coupling to a variety of higher order modes with different azimuthal symmetries.

When the propagating light reaches near the center/waist of the tapered section, it sees a cladding-air waveguide because here the core is insignificant as its size is much smaller than the wavelength of light. The number of modes that this region can support depends on the size of the cladding. As the radius of the taper's waist decreases, the number of modes supported by it decreases. Higher order modes can be filtered out at the waist by making it so thin that it can support only a single mode. This cut-off waist size can be calculated theoretically as briefly explained below, following the detailed treatment in Ref. \onlinecite{Cai,ofm}.

In a step-index, cylindrical waveguide, the electric or magnetic field distribution of the guided modes can be found by solving Maxwell's equations for the boundary conditions imposed by the cylindrical dielectric core (here silica) and cladding (here air). The complete solution can be found in many references (see Ref. \onlinecite{snitzer, ofm, Saleh}). The effective propagation constant $\beta_{eff}$ (axial component of the wave vector) of the guided modes can be found by solving the following transcendental equation \cite{snitzer, ofm, Saleh} 

\begin{equation}
\begin{gathered}
\left( \frac{{J_l}'(pa)}{paJ_l(pa)} + \frac{{K_l}'(qa)}{qaK_l(qa)} \right)\left( \frac{{n_s}^2{J_l}'(pa)}{paJ_l(pa)} + \frac{{n_a}^2{K_l}'(qa)}{qaK_l(qa)} \right)\\
= l^2 \left[\frac{1}{(qa)^2}+\frac{1}{(pa)^2} \right]^2 \left(\frac{\beta_{eff}}{k_0} \right)^2 , 
\end{gathered}
\end{equation}

\noindent with 

\begin{equation}
q^2=\beta_{eff}^2-{n_a}^2 {k_0}^2 , 
\end{equation}
and
\begin{equation}
p^2={n_s}^2 {k_0}^2-\beta_{eff}^2 ,
\end{equation}

\noindent where $J_l$ and $K_l$ are the modified Bessel functions of first and second kind respectively, $l$ is the mode number which is a non-negative integer, $a$ is the radius of the core (here silica), $k_0$ is the propagation constant in vacuum, $n_s$ and $n_a$ are the refractive indices of silica and air respectively. Here, $n_s=1.45$ and $n_a=1$ and wavelength is 780~nm. This equation was solved for first few fiber modes and their effective index $n_{eff}$ as a function of waist diameter ($2\,a$) is plotted in Fig. \ref{neff_plot}. $n_{eff}$ is given by
\begin{equation}
n_{eff} = \frac{\beta_{eff}}{k_0} .
\end{equation}

In a thick waist, the light is mostly confined inside the silica, leading to an effective index close to $n_s = 1.45$. As the waist size decreases down to sub-wavelength scale, the evanescent field in the air region increases and hence the effective index of the taper's waist decreases from $n_s=1.45$ towards $n_a=1$. \cite{Lin} It is clear from the plot that the last higher order modes ($TE_{01}$ and $TM_{01}$) cut-off at $\approx 570$~nm waist diameter. So tapered fibers, with waist diameters less than $570$~nm, can filter out all the higher order modes at $780$~nm or higher wavelengths. The optical microscope (with 1~{\textmu}m resolution) in the set up can't resolve such small waist diameters. Therefore, we observe the transmission through the fiber while tapering to identify the stopping point at which all the coupled higher order modes cut-off. This is explained in the next section.

\section{\label{sec:level5}ANALYSIS OF TRANSMISSION SIGNAL}

\begin{figure}[t]
\centering
\includegraphics[width=1.0\linewidth]{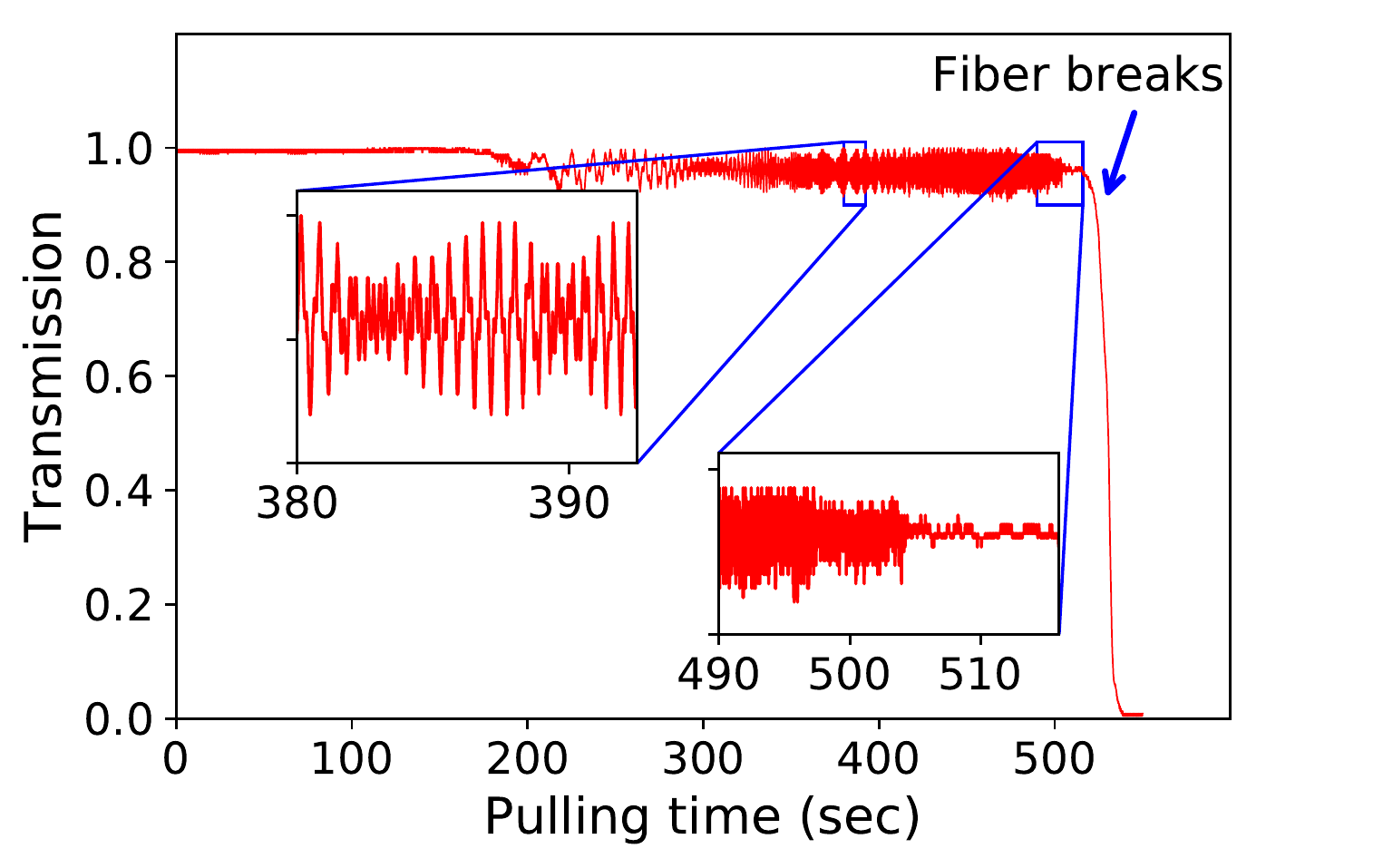} 
\caption{Transmission through a fiber while tapering as a function of pulling time. \textbf{Left inset :} A zoomed view of the oscillations and beating pattern, showing interference between the fiber modes. \textbf{Right inset :} Zoom on the transition of tapered fiber from multi-mode to single mode (significant drop in the amplitude of oscillations).}
\label{transmission_plot}
\end{figure}

As mentioned before, the power transmission through the fibers is observed while heating and pulling the fibers. The changes in output transmission with pulling time for a nearly adiabatic tapered fiber is shown in Fig. \ref{transmission_plot}. The changes in the output transmission result from the interference of different modes supported by the tapered fiber. \cite{Lin, Ravets, Hauer} Different modes propagate with different propagation constants and hence recombine at the output with different acquired relative phases (with respect to the phase of fundamental mode), depending on length and thickness of the tapered fiber. The interference signal oscillates because the acquired relative phases change with pulling time. Oscillations have different frequency components because the phases acquired by different modes change with different rates, causing beating patterns. The amplitude of these oscillations is a measure of the efficiency of excitation and recombination of these modes, which change with pulling time. The component frequencies also change with pulling time because the rates of change of phases are time-dependent during pulling. A zoomed view of the oscillations and the beating pattern is shown in the inset plot (on the left) inside Fig. \ref{transmission_plot}.

\begin{figure}[t]
\centering
\includegraphics[width=0.9\linewidth]{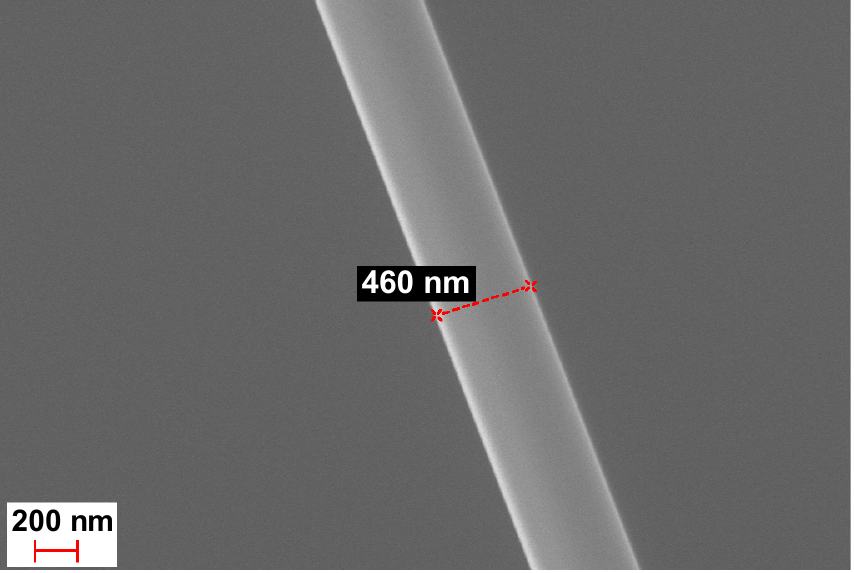} 
\caption{SEM image of a tapered fiber's waist.}
\label{SEM_image}
\end{figure}

In the plot, the point at which the amplitude of the oscillations dies down is when the fiber waist becomes single mode. A zoom on the single mode regime is shown in the inset plot (on the right) inside Fig. \ref{transmission_plot}. The heating and pulling process is stopped immediately after this point is reached. If the pulling continues, then the fiber breaks after 10 to 20 seconds of this single mode operating point (which correspond to just 400 to 800 microns extra pulling length). As estimated in the previous section, the last higher order modes ($TE_{01}$ and $TM_{01}$) cut-off at $\approx 570$~nm waist diameter. Therefore, if $TE_{01}$ or $TM_{01}$ modes get coupled along with other higher modes, while the fundamental mode is propagating down the tapered slope, then the single mode operating point is reached at nearly $570$~nm waist diameter. However, if these modes don't get coupled while other higher modes get coupled, then the single mode operating point is reached before $570$~nm waist diameter. The complete analysis has been done in Ref. \onlinecite{Ravets, Orucevic}. The experimental cut-off waist sizes (at which the last ``coupled'' higher mode cuts off) of the tapered fibers fabricated in our set up are consistent with the theoretical estimation. Waist sizes of most of the tapered fibers were below $570$~nm. An SEM image of a tapered fiber's waist is shown in Fig. \ref{SEM_image}. Its waist size is $460$~nm. The little extra pulling time, during which the fibers continue to get tapered even after the single mode operating point is reached, results in reducing the final waist size below the cut-off waist size.

The tapered fibers fabricated in our set up have greater than $85\%$ transmission on average and the best ones go above $95\%$. We found that cleaning the fiber using hydrogen-oxygen flame, before starting the pulling process, significantly reduces the chances of bad tapering (i.e. high power loss and more coupling to higher order modes). 

\section{\label{sec:level6}BIREFRINGENCE AND CROSS-TALK IN A \mbox{TAPERED} FIBER}

It is important to characterize a tapered fiber for its polarization response, to ensure quantitative accuracy in experiments and measurements. In normal fibers, small amounts of random birefringence, stress and bends cause cross-talk and phase differences between the two polarization modes (say vertical and horizontal). Therefore, the polarization state of the output light is different from the polarization state of the input light. Every fixed polarization state can be described with its two parameters, ellipticity/eccentricity and tilt/orientation angle. Here, ellipticity is defined  as
\begin{equation}
e=\sqrt{1-\frac{E_{min}^2}{E_{max}^2}}=\sqrt{1-\frac{P_{min} }{P_{max}}},
\label{ellipticity}
\end{equation}
where $E_{min}$ and $E_{max}$ are the amplitudes of the maximum and minimum components of the electric field respectively and $P_{min}$ and $P_{max}$ are the measured power values corresponding to $E_{min}$ and $E_{max}$ respectively. Ellipticity has values from $0$ (for circularly polarized light) to $1$ (for linearly polarized light). 

\begin{figure}[t]
 \centering
\includegraphics[width=1.0\linewidth]{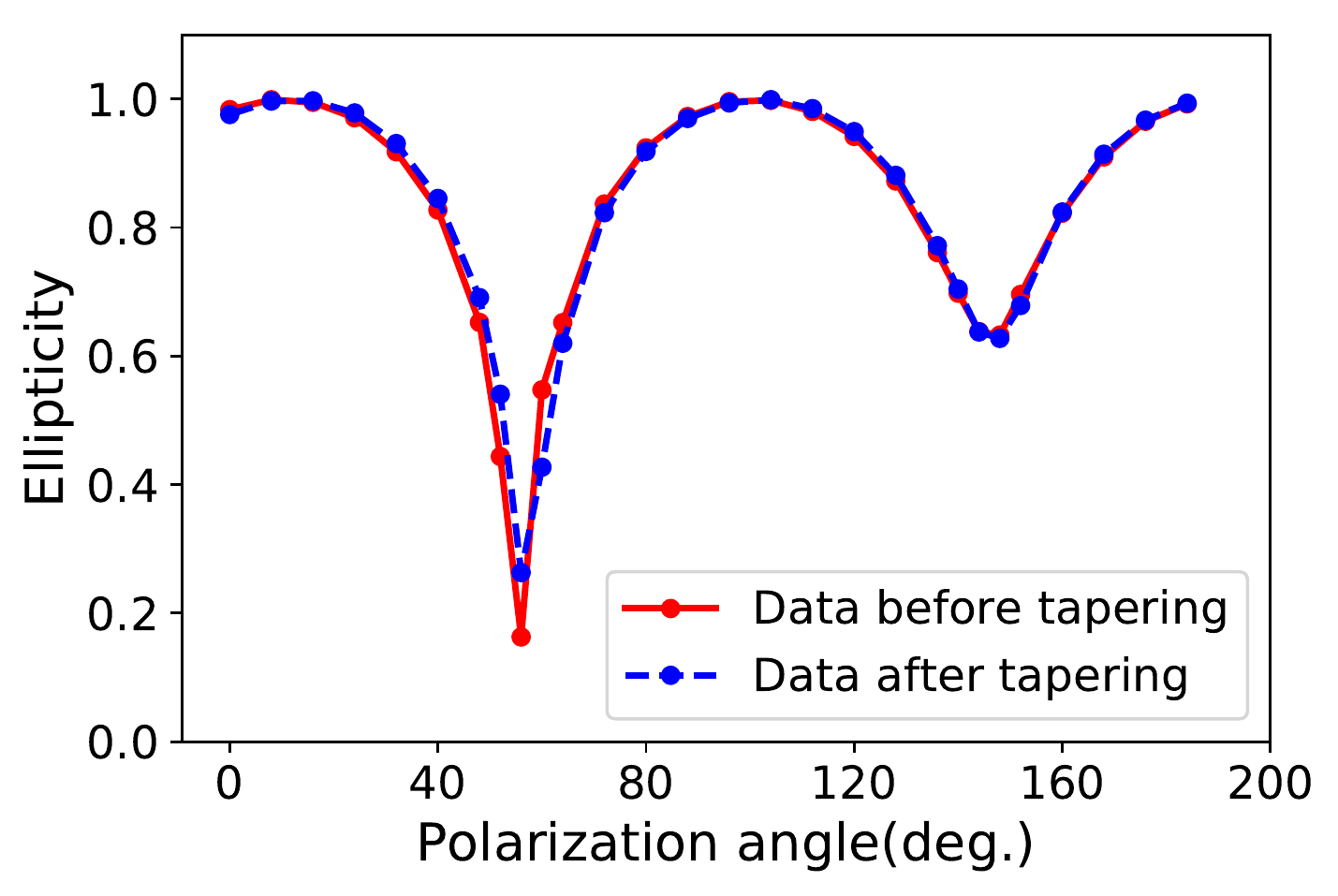} 
\caption{Ellipticity of the transmitted light Vs polarization angle of the input linearly polarized light for an optical fiber before and after tapering. The red and the blue lines are guides to  the eyes. }
\label{pol}
\end{figure}

\vspace{2mm}
Fig. \ref{pol} shows a plot for ellipticity of the transmitted light vs polarization angle (with respect to a reference polarization angle) of the input light before and after tapering of an optical fiber. For this measurement, a linearly polarized laser beam of 780 nm wavelength was coupled to a single mode bare fiber clamped on the heat and pull set up (see figure \ref{ftfs}). The polarization angle of the input light was rotated using a half waveplate before the fiber coupler. The waveplate was rotated manually in steps of 2 degree which rotates the polarization angle by 4 degree. At the output, a polarizer was kept before the detector and it was rotated by 360 degree to measure the power components at all the angles. Out of all the power components, the maximum ($P_{max}$) and minimum ($P_{min}$) components were selected for calculating ellipticity. The piezo-driven polarizer rotator mount takes approximately 135 seconds for one complete rotation. Then the same measurements were taken after tapering the fiber. The plotted data, shown in Fig. \ref{pol}, is for a tapered fiber with transmission $>80\%$. 

As explained in sections 4 and 5, fibers are tapered down to the waist size at which all the coupled higher order modes filter out. The only guided mode that lefts out can be defined as a combination of two orthogonal, quasi-linearly polarized fundamental modes $HE^x_{11}$ and $HE^y_{11}$.\cite{Lei, Kien} In the plot (Fig. \ref{pol}), the ellipticity values are nearly the same before and after tapering. This shows that negligible birefringence and cross-talks between the two polarization modes were caused due to tapering, which further implies that the circular symmetry of the fiber was maintained intact during tapering. While doing this experiment, it was ensured that the conditions which cause birefringence and cross-talk between the two polarization modes (like bends and loops in the fiber, pressure due to the fiber clamps, etc.) remain the same before and after tapering the fiber. Also, we verified that the ellipticity changes by $\leq 0.05$ due to any stretching of the fiber from pulling with motorized stages, or due to long term drifts (1 to 24 hrs observations).

The polarization at the output of a single mode optical fiber as a function of the input polarization has been theoretically modeled in Ref. \onlinecite{Chartier}, along with its experimental verification. In this paper, the optical fiber is modeled as a series of concatenated linear birefringent waveplates, with different amounts of phase retardation and different orientations,  neglecting any polarization-dependent losses. They used the ellipticity angle $\chi$ to represent ellipticity, which is related to the definition of ellipticity used by us by the relation \ $e = \sqrt{1-\tan^2 \chi}$. Their theoretical model explains the general shape of the curve in Fig. \ref{pol}. The dissimilarity between the two minima at $56^\circ$ and $148^\circ$  was already present in the fiber used, before tapering. The asymmetry might be linked to the details of the splicing between the bare fiber and the patch cable, which is necessary for launching light in the coupling fiber. In any case, we verified in this experiment that tapering down the fiber to sub-micron scale did not change its polarization response, as shown in Fig. \ref{pol} 

We also checked, with a tapered fiber with transmission $>90\%$, if the dependence of its transmission on the input polarization changes due to tapering (or if any polarization-dependent losses were introduced due to tapering). As expected, no change was found due to tapering.

\section{\label{sec:level7}\textit{Q}-VALUE MEASUREMENTS OF \texorpdfstring{WGM\MakeLowercase{s}}{WGMs} IN SOLID MICROSPHERES}

\begin{figure}[b]
 \centering
\includegraphics[width=1.0\linewidth]{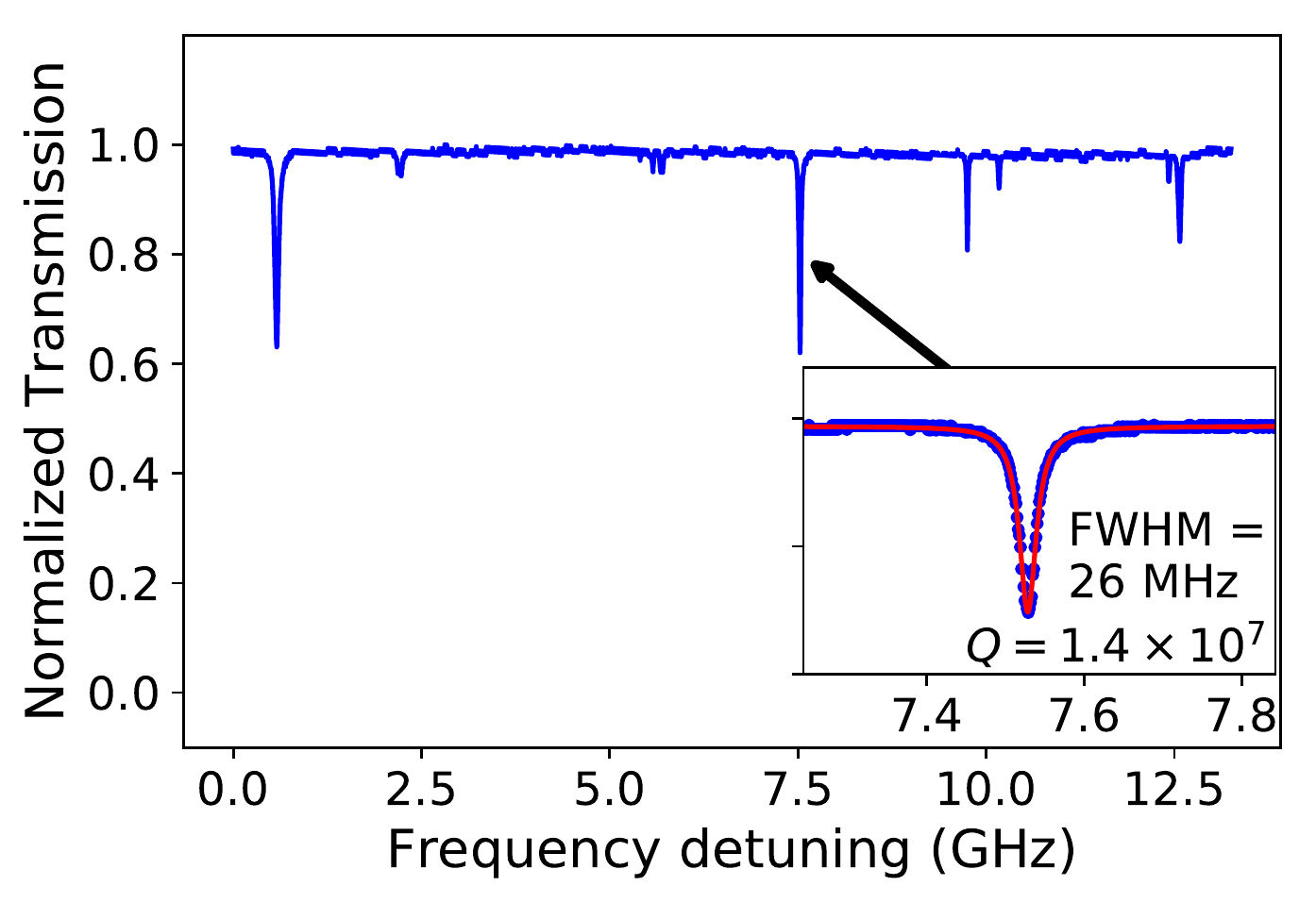} 
\caption{Transmission spectrum of a microsphere-taper coupled system (microsphere diameter $\approx 60$~{\textmu}m). The dips in the signal correspond to the WGMs of the microsphere. The theoretical Lorentzian fit for one of the dips has been shown in the inset plot. In the inset plot, the blue marks are the experiment data points and the red curve is the theoretical fit. }
\label{wgms}
\end{figure}

We have used the tapered fibers for efficiently exciting and detecting high-$Q$ WGMs in silica microspheres of sizes 30 to 300~{\textmu}m. Fig. \ref{mic_tap} shows a microscope image of a silica microsphere kept very close to a tapered fiber to excite its WGMs. To couple light in, a microsphere is moved close to a tapered fiber and a very small gap of sub-wavelength order is maintained between the two to work in the under-coupled regime. In this regime, the observed/loaded $Q$ of the cavity tends to the intrinsic/unloaded $Q$ of the cavity because the external loss due to coupling is minimum and hence the observed $Q$ has its maximum value in the under-coupled regime (see Ref. \onlinecite{Lin}). Fig. \ref{wgms} shows the transmission spectrum of a microsphere and tapered-fiber coupled system. The dips in the signal correspond to the WGMs of the microsphere.  Different dips correspond to different WGMs with different field distributions and angular orientations. Therefore, different modes have different coupling efficiencies depending on the evanescent field distribution and position and orientation of the tapered fiber's waist with respect to the microsphere. The inset shows a Lorentzian function fitted to one of the dips for measuring the Quality factor of the mode. We routinely observe $Q$-factors exceeding $10^7$.

\section{\label{sec:level8}\textit{Q}-VALUE MEASUREMENTS OF \texorpdfstring{WGM\MakeLowercase{s}}{WGMs} IN \mbox{LIQUID} MICRODROPS}

\begin{figure}[b]
 \centering
\includegraphics[width=1.0\linewidth]{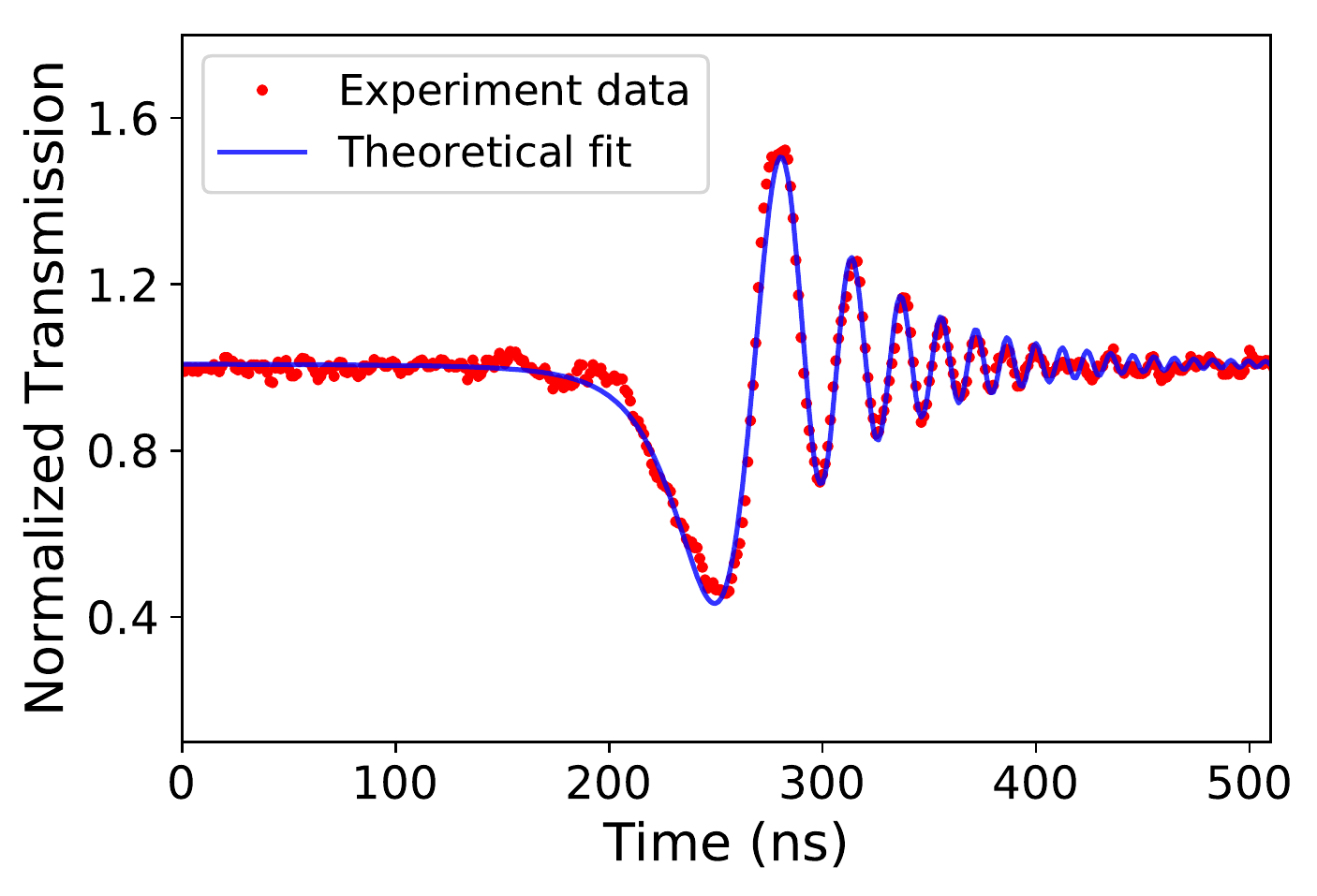} 
\caption{Cavity ring down signal of a WGM in a silicone oil microdrop. Fitted parameters, $V$ (frequency scanning speed) = $4.6 \times 10^5$~ GHz/s, $\tau$ (amplitude lifetime) = 54~ns $\Rightarrow \tau_{photon}$ (photon lifetime) = $\tau/2$ = 27~ns $\Rightarrow$ $Q = 6.5 \times 10^7$}
\label{sil_Q}
\end{figure} 

A microscope image of a silicone oil drop on a silica microsphere is shown in Fig. \ref{sil_drop}. We have done experiments with drops of sizes 50 to 200~{\textmu}m. Unlike the case of solid cavities, in a liquid drop, thermally induced broad-spectrum fluctuations of its size and shape affect the resonance signal. The significantly affecting thermal fluctuations lie in the frequency range from kHz to 100s of kHz. These thermal fluctuations inhibit the reliable measurement of the $Q$ values which are greater than or close to the thermal limits on the measured $Q$ values (see Ref. \onlinecite{Lai, Jonas, Gaira_conf, Giorgini}), when using the standard `linewidth measurement' method. To overcome this, the laser frequency is scanned fast enough to observe the dynamic interference between the direct light propagating through the coupling tapered fiber, the frequency of which is changing very fast, and the out-coupled (stored) resonant light from the high Q microdrop cavity. For observing dynamic interference (or cavity ring-down), the laser frequency has to be scanned fast enough such that the cavity comes and goes out of resonance in duration less than or close to the decay time of the mode. We use $10^5$ to $10^6$~GHz/s scanning speeds for observing the cavity ring-down signals with microdrops of silicone oils. The ring-down data for a WGM resonance in a silicone oil drop has been plotted in Fig. \ref{sil_Q}. The analytical expression for the cavity ring-down curve of a taper-fiber coupled micro-cavity is given in Ref. \onlinecite{Rasoloniaina, Dumeige, Ye}. It has been used for theoretically fitting the experimental data. The fitting gives a $Q$-value of $6.5 \times 10^7$.  

The observed signal can be explained as follows. As the laser frequency and the resonance frequency match, the mode builds up in the cavity. Immediately after that, the laser frequency detunes from the resonance frequency so fast that we can see the exponential decay of the light stored in the cavity. The light which is coupling out of the cavity interferes with the laser light whose frequency is drifting very fast. This results in oscillations in the transmission signal along with an exponential envelope.

We have made $Q$-measurements for drops of different silicone oils.\cite{Gaira_conf} The plotted data in Fig \ref{sil_Q} is with a drop of Sigma-Aldrich's product no. 378364 (polydimethylsiloxane with 100~cSt viscosity at $25^\circ C$ ).

\section{\label{sec:level9}SUMMARY}

We have described an integrated table-top facility for the study of high-$Q$ WGMs in solid and liquid micro-cavities. The facility is self-contained and dust-protected with the modular setups for fabrication of micro-cavities and tapered fibers, and for exciting WGMs in solid and liquid micro-cavities, which are all built on a single floated optical table and inside a custom-designed laminar flow assembly. All the required fabrication and experiments can be done within a small cleanroom-like chamber without the need of transferring the sensitive and fragile elements from one place to another. We have explained in detail the fabrication and performance analysis of tapered fibers of sub-micron waist sizes. The experimental parameters and precautions that are important for fabricating good quality tapered fibers have been explained. The adiabaticity of the fibers can be checked while tapering by observing the transmission signal of the fiber. This also helps in identifying the stopping point when all the coupled higher order modes cut-off. These cut-off waist diameters of the tapered fibers are consistent with the theoretical estimation. The fabricated tapered fibers have greater than 85\% transmission on average and they can completely filter out the unwanted higher order modes. These tapered fibers have been successfully used for efficiently exciting WGMs of $Q\sim10^7$ in silica microspheres and of $Q\  \text{up to}\ 7 \times 10^7$ in microdrops of silicone oils. The $Q$-values of these WGM resonances were measured using `linewidth measurement' for silica microspheres and using `cavity ring-down' technique for silicone oils' microdrops. The capability to track and measure the resonances in microdrops with a wide range of high Q values promises many interesting studies on these dynamic cavities, using our integrated experimental facility.

\section{ACKNOWLEDGEMENTS}
We are thankful to P.G. Rodrigues, P.V. Sudersanan, S.K. Guram and the central workshop facility in TIFR for the design and fabrication of many components and units of the setup. M. Gaira thanks S. Selvarajan, N. Jetty and D.B. Dhuri for useful discussions.

\end{document}